\documentclass[conference]{IEEEtran}
\IEEEoverridecommandlockouts

\usepackage{cite}
\usepackage{bm}
\usepackage{amsmath,amssymb,amsfonts}
\usepackage{algorithmic}
\usepackage{graphicx}
\usepackage{textcomp}
\usepackage{xcolor}
\usepackage{xspace}
\usepackage{url}
\usepackage{tabularx, booktabs,}

\def\res{\texttt{ResUNet}\xspace}
\def\resk{\texttt{ResUNetK}\xspace}

\def\mla{\texttt{CA}\xspace}
\def\mlatp{\texttt{CA-TP}\xspace}
\def\mpipad{\texttt{PI-Pad}\xspace}
\def\mpipen{\texttt{PI-Pen}\xspace}

\def\BibTeX{{\rm B\kern-.05em{\sc i\kern-.025em b}\kern-.08em
    T\kern-.1667em\lower.7ex\hbox{E}\kern-.125emX}}
\begin{document}

\title{Baseline Systems and Evaluation Metrics for Spatial Semantic Segmentation of Sound Scenes\\
\thanks{This work was partially supported by JST Strategic International Collaborative Research Program (SICORP), Grant Number JPMJSC2306, Japan.

* These authors contributed equally to this work.}
}

\author{\IEEEauthorblockN{Binh Thien Nguyen*}
\IEEEauthorblockA{
\textit{NTT Corporation}\\
Japan
}
\and
\IEEEauthorblockN{Masahiro Yasuda*}
\IEEEauthorblockA{
\textit{NTT Corporation}\\
Japan 
}
\and
\IEEEauthorblockN{Daiki Takeuchi}
\IEEEauthorblockA{
\textit{NTT Corporation}\\
Japan
}
\and
\IEEEauthorblockN{Daisuke Niizumi}
\IEEEauthorblockA{
\textit{NTT Corporation}\\
Japan 
}
\and
\IEEEauthorblockN{Yasunori Ohishi}
\IEEEauthorblockA{
\textit{NTT Corporation}\\
Japan 
}
\and
\IEEEauthorblockN{Noboru Harada}
\IEEEauthorblockA{
\textit{NTT Corporation}\\
Japan 
}
}

\maketitle

\begin{abstract}

Immersive communication has made significant advancements, especially with the release of the codec for Immersive Voice and Audio Services.
Aiming at its further realization, the DCASE 2025 Challenge has recently introduced a task for spatial semantic segmentation of sound scenes (S5), which focuses on detecting and separating sound events in spatial sound scenes.
In this paper, we explore methods for addressing the S5 task.
Specifically, we present baseline S5 systems that combine audio tagging (AT) and label-queried source separation (LSS) models.
We investigate two LSS approaches based on the ResUNet architecture: a) extracting a single source for each detected event and b) querying multiple sources concurrently.
Since each separated source in S5 is identified by its sound event class label, we propose new class-aware metrics to evaluate both the sound sources and labels simultaneously.
Experimental results on first-order ambisonics spatial audio demonstrate the effectiveness of the proposed systems and confirm the efficacy of the metrics.

\end{abstract}



\section{Introduction} \label{sec:intro}
Recently, immersive communication has emerged as a promising field and gained attention \cite{ivas, ios, ivas_app}, especially with the release of a codec for Immersive Voice and Audio Services (IVAS) \cite{ivas}.
Fundamental technologies for realizing immersive communication include decomposing a complex spatial sound scene into sound sources, accompanied by metadata describing their properties, such as the direction of arrival.
These technologies are essential for spatial audio formats such as object-based audio \cite{objectbased} and Metadata-Assisted Spatial Audio (MASA) \cite{masa}.
In these formats, spatial audio can be coded with fewer channels, reducing complexity and transmission delay, making them especially suitable for immersive communication on low computational capacity devices like smartphones.

Aiming for immersive communication, the DCASE 2025 Challenge has recently introduced a task on sound event detection and separation from spatial sound scenes, referred to as ``Task 4: Spatial semantic segmentation of sound scenes (S5)\footnote{https://dcase.community/challenge2025/\#task4}."
An S5 system takes as input a reverberant multi-channel spatial audio mixture, consisting of multiple sound sources and diffuse background noise, and outputs the isolated single-channel dry sound sources along with their respective sound event class labels.
This task adds complementary functions to other studies that estimate other meta-information, such as \cite{seld}, contributing to the realization of immersive communication.

Separating sound mixtures into sources, known as source separation (SS), and predicting audio class labels, referred to as audio tagging (AT) or sound event detection (SED), are active areas of research, with their combination also being explored.
SS has been used as a preprocessing step to improve the results of SED \cite{uss_sed}, and this approach served as the baseline for Task 4 of the DCASE Challenge 2020 and 2021.
In contrast, \cite{sed_uss} and \cite{m2dsb} show that the performance of an SS network can be enhanced by incorporating the embedding information extracted by a sound classifier model.
Further to this, \cite{uss} employs SED and AT to isolate individual sound sources from a weakly labeled signal, facilitating the training of a label-queried SS (LSS) model on a large-scale dataset.
As another approach, \cite{deftmamba} utilizes a multi-task learning framework for both source separation and label prediction.
However, the main goal of these works is to improve either or both the SS and AT, while their performances are evaluated separately, or only the performance of the main task is considered.
The connection between the separated sources and the predicted labels has not been thoroughly evaluated, which contrasts with the S5 task, where a separated source is identified by its label.

In this paper, we explore approaches for the S5 task.
Specifically, we introduce baseline systems combining AT and LSS models, built upon a fine-tuned Masked Modeling Duo (M2D) model \cite{m2d} and a modified version of ResUNet \cite{uss}.
For LSS, in addition to extracting a single source as in \cite{uss}, we also explore a model that queries multiple sound sources simultaneously.
Furthermore, we propose new class-aware metrics that evaluate separated sources and their corresponding labels simultaneously.
The proposed task setting, metrics, and systems will be reflected in the DCASE 2025 Challenge Task~4.
The code is publicly available on GitHub\footnote{\url{https://github.com/nttcslab/dcase2025_task4_baseline}}.

\section{Related Work} \label{sec:related_work}
\subsection{M2D-based Audio Tagging}
An M2D-based AT model is a fine-tuned M2D \cite{m2d} model on AudioSet \cite{audioset} with 527 classes. M2D is a self-supervised learning (SSL) foundation model that is pre-trained with M2D's masked prediction-based objective using only audio samples from AudioSet.
It has been highlighted that the audio representations extracted using M2D achieve state-of-the-art performance in various downstream tasks in diverse domains such as speech, music, and environmental sound.
Furthermore, M2D-X \cite{m2d}, an extension of M2D, accommodates additional training objectives for learning representations specialized to applications. M2D-X provides a variant, M2D-AS, tailored to AudioSet tagging using a combination of SSL objective and additional supervised loss for label prediction.
We utilize an M2D-based AT model with an M2D-AS backbone for label prediction in our system.

\subsection{Label-Queried Source Separation Using ResUNet}
Kong et al. \cite{uss} proposed a single-input, single-output model for LSS.
The model takes as input a single-channel sound mixture and a one-hot vector representing the sound class label, and it outputs the separated target sound source corresponding to the label query.
The input signal is first converted into magnitude and phase spectrograms using the short-time Fourier transform (STFT).
Then, the magnitude spectrogram is processed by an encoder-decoder separator network, built on a ResUNet backbone.
The separator network consists of 6 symmetric encoder and decoder blocks with skip-connections, where each block consists of 2D convolutional/deconvolutional layers, batch normalization, and leaky ReLU activation layers.
The label query information is incorporated into the separator network using Feature-wise Linear Modulation (FiLM) \cite{film}.
The separator network outputs the magnitude mask and phase residual, which are applied to the magnitude and phase spectrograms of the input mixture to obtain the spectrograms of the target sound source.
Finally, the target sound source is reconstructed using the inverse STFT.
Experimental results \cite{uss} have demonstrated the effectiveness of LSS in a wide range of separation tasks, including music and general sound separation.

\section{Proposed Spatial Semantic Segmentation Systems}
\subsection{Task Setting}
Let $\bm{Y} = [\bm{y}^{(1)},\dots, \bm{y}^{(M)}]^\top \in \mathbb{R}^{M \times T}$ be the multi-channel time-domain mixture signal of length $T$, recorded with an array of $M$ microphones, where $\{\cdot\}^\top$ is the matrix transposition.
We denote $C=\{c_1, ...,c_K\}$ the set of source labels in the mixture, where the source count $K$ can vary from $1$ to $K_\textrm{max}$.
The $m$-th channel of $\bm{Y}$ can be modeled as
\begin{equation}\label{eq:mixture}
	\bm{y}^{(m)}=\sum_{k=1}^{K} \bm{h}^{(m)}_k*\bm{s}_k + \bm{n}^{(m)} 
	            =\sum_{k=1}^{K}\bm{x}^{(m)}_k + \bm{n}^{(m)},
\end{equation}
where $\bm{s}_k\in\mathbb{R}^T$ is the single-channel dry source signal corresponding to the label $c_k$, $\bm{h}^{(m)}_k \in \mathbb{R}^H$ is the $m$-th channel of the length-$H$ room impulse response (RIR) at the spatial position of $\bm{s}_k$, and $\bm{n}^{(m)}\in\mathbb{R}^T$ is the $m$-th channel of the multi-channel noise signal.
The wet source $\bm{x}^{(m)}_k\in\mathbb{R}^T$ can be split into two components: the direct path and early reflections, $\bm{x}^{(m,\textrm{d})}_k\in\mathbb{R}^T$, and late reverberation, $\bm{x}^{(m,\textrm{r})}_k\in\mathbb{R}^T$, as
\begin{equation}
	\bm{x}^{(m)}_k = \bm{x}^{(m,\textrm{d})}_k + \bm{x}^{(m,\textrm{r})}_k
		               = \bm{h}^{(m,\textrm{d})}_k*\bm{s}_k + \bm{h}^{(m,\textrm{r})}_k*\bm{s}_k,
\end{equation}
where $\bm{h}^{(m,\textrm{d})}_k, \bm{h}^{(m,\textrm{r})}_k \in \mathbb{R}^H$ are the corresponding early and late parts of the RIR, respectively. We denote by $N$ the number of sound event classes.

The goal of S5 is to extract all the single-channel sources $\{\hat{\bm{x}}^{(m_\textrm{ref},\textrm{d})}_1, \dots,\hat{\bm{x}}^{(m_\textrm{ref},\textrm{d})}_{\hat{K}} \}$ at a reference microphone $m_\textrm{ref}$ and their corresponding class labels $\hat{C}=\{\hat{c}_1, \dots,\hat{c}_{\hat{K}}\}$ from the multi-channel mixture $\bm{Y}$.

From now on, we drop some indices for the clarity of the formulation. The notations of $\bm{x}^{(m_\textrm{ref}, \textrm{d})}_k$, $\hat{\bm{x}}^{(m_\textrm{ref}, \textrm{d})}_k$, and $\bm{y}^{(m_\textrm{ref})}$ become $\bm{x}_k$, $\hat{\bm{x}}_k$, and $\bm{y}$, respectively.

\subsection{Proposed Evaluation Metrics}
In this section, we introduce class-aware metrics that evaluate both the sound sources and their class labels simultaneously for the S5 task.
The main idea is that the estimated and reference sources are aligned by their labels, with the waveform metric being calculated when the label is correctly predicted, i.e., $c_k \in C \cap \hat{C}$.
In cases of incorrect label prediction, penalty values are accumulated.
We define a class-aware signal-to-distortion ratio improvement (CA-SDRi) metric as
\begin{equation}
\begin{split}
	\textrm{CA-SDRi}&\left(\{\hat{\bm{x}}_1, \dots,\hat{\bm{x}}_{\hat{K}}\}, \{\bm{x}_1, \dots,\bm{x}_K\}, \hat{C}, C, \bm{y}\right) \\ &= \frac{1}{| C \cup \hat{C} |} \sum_{c_k \in C \cup \hat{C}} P_{c_k},
\end{split}
\end{equation}
where $| C \cup \hat{C} |$ is the length of the set union. The metric component $P_{c_k}$ is calculated as
\begin{equation}
P_{c_k\in C \cup \hat{C}} = 
\begin{cases}
	\textrm{SDRi}(\hat{\bm{x}}_k, \bm{x}_k, \bm{y}), &\text{if } c_k \in C \cap \hat{C}\\
	\mathcal{P}^\textrm{FN}_{c_k}, &\text{if } c_k \in C \text{ and } c_k \notin \hat{C}\\
	\mathcal{P}^\textrm{FP}_{c_k}, &\text{if } c_k \notin C \text{ and } c_k \in \hat{C}
\end{cases},
\end{equation}
where the SDRi is calculated as
\begin{equation}
	\textrm{SDRi}(\hat{\bm{x}}_k, \bm{x}_k, \bm{y})
	= \textrm{SDR}(\hat{\bm{x}}_k, \bm{x}_k) - \textrm{SDR}(\bm{y} , \bm{x}_k),
\end{equation}
\begin{equation}\label{sq:sdr}
	\textrm{SDR}(\hat{\bm{x}}, \bm{x}) 
	= 10\log_{10} \left( \frac{\|\bm{x}\|^2}{\|\bm{x} - \hat{\bm{x}}\|^2} \right).
\end{equation}
$\mathcal{P}^\textrm{FN}_{c_k}$ and $\mathcal{P}^\textrm{FP}_{c_k}$ are the penalty values for false negative (FN) and false positive (FP), both set to $0$, indicating that incorrect predictions do not contribute to any improvement in the metric.

In a similar calculation method, the class-aware scale-invariant SDRi (CA-SI-SDRi), can be calculated by replacing the SDR in (\ref{sq:sdr}) with the SI-SDR \cite{sisdr} defined as
\begin{equation}
	\textrm{SI-SDR}(\hat{\bm{x}}, \bm{x}) 
	= 10\log_{10} \left( \frac{ \| \alpha \bm{x}\|^2}{\| \alpha \bm{x} - \hat{\bm{x}}\|^2} \right),
\end{equation}
where $\alpha = \hat{\bm{x}}^\top \bm{x}/\|\bm{x}\|^2$ is a scaling factor.


\subsubsection{Relation to Conventional Metrics} \label{ssec:relation_metrics}
Our metrics share some similarities with conventional metrics \cite{metric_pad0, metric_pad0_eps, metric_pen} for source separation with an unknown number of sources, where the mixture contains $K$ sources and the model outputs $\hat{K}$ sources.
When $\hat{K}=K$, these metrics are typically computed using a permutation-invariant objective, where the output sources are permuted to best match the reference sources in order to maximize the metrics.
When $\hat{K}\neq K$, one approach \cite{metric_pad0, metric_pad0_eps} is to pad all-zero signals to the estimated sources when the source count is underestimated ($\hat{K}<K$) and to retain only a subset that best matches the reference sources in the case of overestimation ($\hat{K} > K$).
Alternatively, \cite{metric_pen} proposed a penalized metric composed of two components: the waveform metrics calculated on the best-matching subsets of size $\min(K, \hat{K})$ from the estimated and reference sources, and the penalties for source count deviation, $|K - \hat{K}|\cdot\mathcal{P}_\textrm{ref}$, where $\mathcal{P}_\textrm{ref}$ is a fixed penalty value.

The primary distinction is that our metric identifies sources using their labels, while these conventional metrics align sources using a permutation-invariant objective.
In addition, unlike \cite{metric_pen}, which calculates the penalty based on source count deviation, our metric explicitly penalizes the FP and FN cases, avoiding situations where the source count is correct but the label prediction is incorrect.

\subsection{Proposed Systems}\label{ss:baseline}
\begin{figure}[t]
\begin{center}
\includegraphics[trim={0 0 0 0},clip,scale=1]{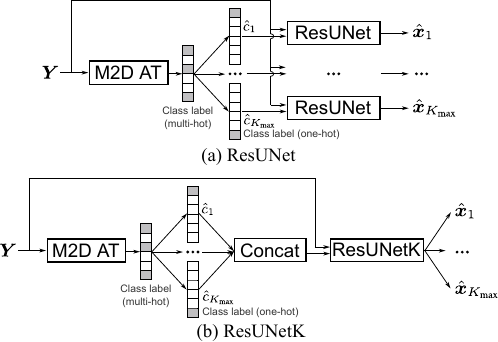}
\end{center}
\vspace{-0.3cm}
\caption{Proposed spatial semantic segmentation system flow.}\label{fig:baseline}
\vspace{-0.1cm}
\end{figure}
To address the S5 task, we introduce two two-stage systems, (a) ResUNet and (b) ResUNetK, as shown in Fig.~\ref{fig:baseline}.
The first stage predicts the source labels from the mixture, while the second stage extracts the corresponding sound sources.
The details of these systems are as follows.

The first stage is identical for both systems, where we fine-tune an AT model consisting of a feature extractor backbone (i.e., M2D) and a head layer.
The input of the model is the first channel of $\bm{Y}$.
We replace the head of the original M2D model with a linear layer, which outputs the probabilities for $N$ sound event classes in our dataset.
To fine-tune the M2D-based AT, we initially trained only the head while keeping the rest of the model frozen. Next, we unfroze the last two layers and fine-tuned them along with the head for several epochs.
Binary cross-entropy was used as the loss function.
After fine-tuning, $\hat{C}$ is obtained by applying a threshold $\gamma$, with the estimated number of sources, $\hat{K}$, limited to the range $[1, K_\textrm{max}]$.
If $\hat{K} < 1$, the highest probability is selected.
If $\hat{K} > K_\textrm{max}$, the top $K_\textrm{max}$ probabilities are chosen.
It is worth noting that the output of the first stage in Fig.~\ref{fig:baseline} consists of $K_\textrm{max}$ label vectors, and if $\hat{K} < K_\textrm{max}$, the remaining $(K_\textrm{max} - \hat{K})$ vectors are all zeros.
These zero labels, however, are not considered in the set $\hat{C}$ during the evaluation phase. 

For the second stage of the first system in Fig.~\ref{fig:baseline}(a), we modify the single-input, single-output ResUNet from \cite{uss}.
We adjust the input layer to accept the $M$-channel mixture spectrogram and modify the output layer to predict the $M$-channel magnitude mask and phase residual, which are applied to the input spectrogram.
Following this, we add a $1 \times 1$ convolutional layer to generate the single-channel spectrogram for the target source.
During inference, ResUNet is executed $\hat{K}$ times to extract $\hat{K}$ sources from the mixture.

Furthermore, we introduce a single-input, multiple-output variant of ResUNet, named ResUNetK, as illustrated in Fig.~\ref{fig:baseline}(b).
ResUNetK has a similar structure to ResUNet, except that it outputs $K_\textrm{max} \times M$-channel magnitude mask and phase residual.
These mask and residual are applied to the input mixture spectrogram, followed by a $1 \times 1$ convolutional layer that generates $K_\textrm{max}$ single-channel spectrograms, corresponding to $K_\textrm{max}$ output sound sources.
Another difference is that the query of ResUNetK is a vector of length $K_\textrm{max} \times N$, which is a concatenation of $K_\textrm{max}$ one-hot label vectors.
The output sound sources are generated in the same order as the label vectors in the query.
In addition, we add a connection so that if the label contains only zeros, the corresponding output waveform will also be all zeros.

Both the ResUNet and ResUNetK are trained with reference labels and sound sources using the SDR loss function.
For ResUNetK, the loss is the mean SDR of all output sources. We randomly shuffled the labels to ensure that the model could extract the sources regardless of the label order.

\section{Experiments and Results} \label{sec:experiments}
\subsection{Experimental Setting}
For the test data, we generated $1296$ mixtures from the test split of sound sources and RIR datasets, with $432$ mixtures containing $1$, $2$, and $3$ sources, respectively.
For validation, a similar number of mixtures was generated from the validation split, while for training, mixtures were synthesized on-the-fly.

All sources were sampled at $32$ kHz with the duration $T$ of $10$ seconds.
Each mixture may contain $1$ to $K_\textrm{max}=3$ sources.

For the sound sources $\bm{s}_k$, we used the same train, validation, and test splits of the dataset in \cite{semhear}, which is derived from FSD50K \cite{fsd50k} and ESC-50 \cite{esc50}.
We selected only 18 out of the 20 available sound classes, excluding ``music" and ``singing".

For the RIR and noise, we used the FOA-MEIR dataset \cite{foameir}, which contains first-order ambisonics RIR recorded in various environments.
The Reverb-C and Test subsets of FOA-MEIR, consisting of 7 environments and 216 IR recordings each, were used for training.
The Reverb-S subset, comprising 96 environments with 3 IR recordings each, was randomly split into 48 environments for testing and the rest for validation.

The multi-channel mixtures were synthesized using a modified version of the SpatialScaper toolkit \cite{spatialscaper}.
For each mixture, we first selected a room from the RIR dataset and a $10$-second segment from the corresponding noise signal. 
Then, we randomly generated a number of sources $K\in[1,K_\textrm{max}]$, after which $K$ positions in the room and $K$ sound events were selected.
In this setting, we assumed that the sound events in a mixture are mutually exclusive.
The mixture was then synthesized using (\ref{eq:mixture}), with the SNR of each sound event ranging from $5$ to $20$ dB.
The direct path IR, $\bm{h}^{(m, \textrm{d})}_k$, was calculated by retaining the range of $[-6, 50]$\,ms around the first peak of $\bm{h}^{(m)}_k$ and setting the rest to zero.
The reference channel, $m_\textrm{ref}$, was set to $1$, corresponding to the omnidirectional channel.

For the first stage of the S5 systems, we used the M2D-based AT model available online\footnote{``M2D-AS fine-tuned on AS2M" on \url{https://github.com/nttcslab/m2d}}.
The probability threshold $\gamma$ was set to $0.5$.
For the second stage, both ResUNet and ResUNetK were trained with a batch size of 5 on 8 RTX 3090 GPUs using the Adam optimizer with a learning rate of $10^{-4}$ for 250K steps.
The training conditions were identical for both models, except that, for ResUNet, we randomly selected one source as the target for each mixture.

\subsection{Audio Tagging Results}
\begin{table}[t!]
\caption{Accuracy (\%) of M2D-based Audio Tagging Models}
\vspace{-8pt}
\label{tab:m2d}
\centering
\begin{tabular}{p{3cm}cccc} 
\toprule
Number of sources & 1 & 2 & 3  & Average\\
\midrule
M2D AT (head) & 91.2 & 71.1 & 54.4  & 72.2\\
M2D AT & 93.1 & 81.7 & 79.4  & 84.7\\
\bottomrule
\end{tabular}
\end{table}
Table~\ref{tab:m2d} presents the accuracy of the M2D AT models on the test dataset. ``M2D AT (head)" refers to the checkpoint from the first step of fine-tuning, where only the head layer is trained. 
Since the separation model is LSS, the accuracy of label prediction in the first stage will directly affect the separated sources in the second stage.

\subsection{Assessment of Evaluation Metrics}
\begin{figure}[t]
\begin{center}
\includegraphics[trim={0 0 0 0},clip,scale=1]{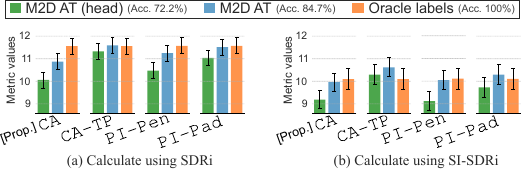}
\end{center}
\vspace{-0.2cm}
\caption{Comparison of various metrics for assessing S5 systems. The bar colors indicate first-stage methods. The second-stage methods are all ResUNet.}\label{fig:metrics}
\end{figure}
To assess the suitability of metrics for S5 systems, we compared our class-aware metrics (\mla) with conventional metrics using permutation-invariant objectives described in Section~\ref{ssec:relation_metrics}, including the penalized metric (\mpipen) \cite{metric_pen} and the zero-padded metric (\mpipad) \cite{metric_pad0_eps}. We set $\mathcal{P}_\textrm{ref}$ in \mpipen to $0$ to be consistent with our metric.
For reference, we also included our metric without penalty (\mlatp), calculated using true positive predictions, with FN and FP ignored.

\begin{figure}[t]
\begin{center}
\includegraphics[trim={0 0 0 0},clip,scale=1]{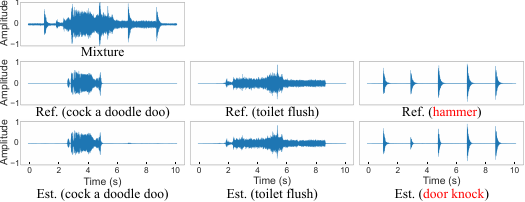}
\end{center}
\vspace{-0.4cm}
\caption{Example of an S5 system when the labels are incorrectly predicted.}\label{fig:waveform_ex2}
\end{figure}

The results are shown in Fig.~\ref{fig:metrics}.
We can see that, with the oracle labels in the first stage, all the metrics yield similar values.
Since the separation model is LSS, given correct labels, if the separated sources are of sufficient quality, aligning the sources by permutations is nearly equivalent to aligning them by labels.
The values vary across the metrics when the labels are predicted.
It appears that \mlatp and \mpipad may not be appropriate metrics, as, when calculated with SI-SDRi, the predicted labels using M2D AT yield even higher metric values than the oracle labels.
\mla and \mpipen are more effective at reflecting the performance of the label prediction, as their metric values increase with higher label prediction accuracy in both SDRi and SI-SDRi cases.
However, for the S5 task, \mpipen may fail to evaluate in certain cases when the labels are incorrectly predicted, an example of which is shown in Fig.~\ref{fig:waveform_ex2}.
In this figure, although the label prediction misclassifies the ``hammer", the source separation model still provides a reasonable estimate of the sound source using the ``door knock" label, which is most likely due to similarities between these two events.
In this case, \mla considers ``hammer" as FN and ``door knock" as FP, while \mpipen, by only evaluating the estimated waveform without considering the label, imposes no penalty.
These observations demonstrate the efficiency of \mla in evaluating the S5 task, compared to conventional metrics.

\begin{figure}[t]
\begin{center}
\includegraphics[trim={0 0 0 0},clip,scale=1]{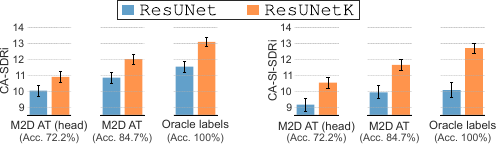}
\end{center}
\vspace{-0.2cm}
\caption{Class-aware metrics on various S5 systems. The x-axis shows first-stage methods, and bar colors indicate second-stage methods.}\label{fig:results}
\end{figure}

\subsection{Evaluation of Proposed Systems using Proposed Metrics}

\begin{figure}[t]
\begin{center}
\includegraphics[trim={0 0 0 0},clip,scale=1]{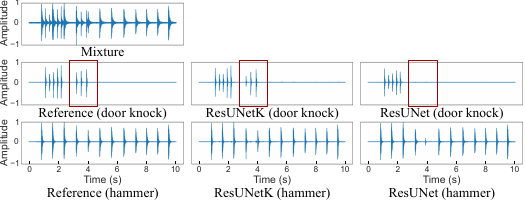}
\end{center}
\vspace{-0.2cm}
\caption{Example of ResUNet and ResUNetK with given labels.}\label{fig:waveform_ex}
\end{figure}

We evaluate the proposed S5 systems: (a) ResUNet (\res) and (b) ResUNetK (\resk), with the proposed evaluation metrics, the results of which are depicted in Fig.~\ref{fig:results}.
It can be seen that \resk outperforms \res in all metrics when the same label predictions are used in the first stage.
A possible reason is that, while being trained under nearly identical conditions, \resk reconstructs all the sources simultaneously, which may facilitate mutual assistance among the reconstructions of these sources.
We observed that in many cases, the sum of the output sources from \resk is close to the mixture, even though we did not impose a loss function for mixture reconstruction.
\res may not benefit from this property since it extracts each source separately.
Fig.~\ref{fig:waveform_ex} shows an example of the superior performance of \resk compared to \res, where the mixture contains two similar sound events, ``door knock" and ``hammer".
As shown, \resk successfully extracted the ``door knock" sound at 4 seconds, while \res did not.
These findings illustrate the advantage of \resk over \res in the S5 task, with \resk not only achieving higher performance but also being faster due to parallel source extractions.

\section{Conclusions} \label{sec_conclusions}
In this paper, we introduce two-stage systems to address the S5 task, aiming at simultaneously detecting and separating sound events from multi-channel mixture signals.
The first stage is an AT model that predicts source labels in the mixture, which is obtained by fine-tuning the pre-trained SSL M2D model.
The second stage is an LSS model that extracts sound sources using the labels predicted in the first stage.
Two variants for the LSS model have been explored: ResUNet, which extracts a single source, and ResUNetK, which simultaneously separates all sources from a mixture. To evaluate the performance of the proposed systems, we also proposed the evaluation metric dedicated to the S5 task. 
Experimental results highlight the effectiveness of the ResUNetK and confirm the efficacy of the proposed metrics in evaluating both source separation and class label prediction simultaneously.

The proposed task setting, metrics, and
systems will be reflected in those of the DCASE 2025 Challenge
Task 4: Spatial Semantic Segmentation of Sound Scenes. We expect to have enhanced solutions for the S5 task with the challenge.


\bibliographystyle{IEEEtran}
\bibliography{ref}
\end{document}